\documentclass[twocolumn,english,superscriptaddress,floatfix]{revtex4}
\usepackage[T1]{fontenc}
\usepackage[latin9]{inputenc}
\usepackage{color}
\definecolor{orange}{cmyk}{0.00, 0.33, 1.00, 0.00}
\definecolor{KIT-green}{RGB}{0, 150,130}
\definecolor{KIT-blue}{RGB}{70,100,170}
\usepackage{babel}
\usepackage{verbatim}
\usepackage{amsmath}
\usepackage{amssymb}
\usepackage{graphicx}
\usepackage{esint}
\usepackage[unicode=true,
 bookmarks=false,
 breaklinks=true,pdfborder={0 0 1},
 colorlinks=true]
 {hyperref}
\hypersetup{
 pdfcreator={},
 pdfproducer={LaTeX with hyperref}, linkcolor=KIT-blue, anchorcolor=KIT-blue, citecolor=KIT-blue, filecolor=red, menucolor=red, pagecolor=red, urlcolor=KIT-blue, pdfstartview=FitV, pdfpagelayout=OneColumn, hypertexnames=true}

\makeatletter

\newcommand{\im}{\text{Im}}

\newcommand{\TRSB}{{\scriptscriptstyle \mathrm{TRSB}}}
\newcommand{\kB}{k_{\scriptscriptstyle B}}
\newcommand{\eF}{\epsilon_{\scriptscriptstyle F}}

\usepackage[figure,table]{hypcap}

\makeatother

\begin{document}
\title{Inhomogeneous time-reversal symmetry breaking in $\mathrm{Sr}_{2}\mathrm{Ru}\mathrm{O}_{4}$}
\author{Roland Willa}
\affiliation{Institute for Theory of Condensed Matter, Karlsruhe Institute of Technology,
Karlsruhe 76131, Germany}
\author{Matthias Hecker }
\affiliation{Institute for Theory of Condensed Matter, Karlsruhe Institute of Technology,
Karlsruhe 76131, Germany}
\author{Rafael M. Fernandes}
\affiliation{School of Physics and Astronomy, University of Minnesota, Minneapolis,
55455 MN} 
\author{Jörg Schmalian}
\affiliation{Institute for Theory of Condensed Matter, Karlsruhe Institute of Technology,
Karlsruhe 76131, Germany}
\affiliation{Institute for Quantum Materials and Technologies, Karlsruhe Institute
of Technology, Karlsruhe 76021, Germany}

\begin{abstract}
We show that the observed time-reversal symmetry breaking (TRSB) of the superconducting state in $\mathrm{Sr}_{2}\mathrm{Ru}\mathrm{O}_{4}$ can be understood as originating from inhomogeneous strain fields near edge dislocations of the crystal. Specifically, we argue that, without strain inhomogeneities, $\mathrm{Sr}_{2}\mathrm{Ru}\mathrm{O}_{4}$ is a single-component, time-reversal symmetric superconductor, likely with $d_{x^{2}-y^{2}}$ symmetry. However, due to the strong strain inhomogeneities generated by dislocations, a slowly-decaying sub-leading pairing state contributes to the condensate in significant portions of the sample. As it phase winds around the dislocation, time-reversal symmetry is locally broken. Global phase locking and TRSB occur at a sharp Ising transition that is not accompanied by a change of the single-particle gap and yields a very small heat capacity anomaly. Our model thus explains the puzzling absence of a measurable heat capacity anomaly at the TRSB transition in strained samples, and the dilute nature of the time-reversal symmetry broken state probed by muon spin rotation experiments. We propose that plastic deformations of the material may be used to manipulate the onset of broken time-reversal symmetry.
\end{abstract}
\maketitle

\section{Introduction}

Establishing the symmetry of the Cooper pair wave function is the
pivotal step to understand a superconductor. It not only determines the
macroscopic phenomenology, but it also narrows down the microscopic mechanism
of the pairing state. Arguably the strongest evidence that cuprate-based
high-temperature superconductors are governed by an electronic mechanism
is the observation of the $d_{x^{2}-y^{2}}$ symmetry of the pair
wave function \cite{Wollman1993,Tsuei1994}. $\mathrm{Sr}_{2}\mathrm{Ru}\mathrm{O}_{4}$ is
a layered perovskite superconductor that is iso-structural to La$_{2}$CuO$_{4}$ \cite{Maeno1994}.
Given the crossover from incoherent to coherent transport as function
of temperature, $\mathrm{Sr}_{2}\mathrm{Ru}\mathrm{O}_{4}$ is---just like the
cuprates---governed by strong electronic correlations \cite{Georges2011}. However,
early on it was advocated that the more appropriate analogue for the
origin and symmetry of the pairing state might be $^{3}$He \cite{Rice1995,Baskaran1996}.
Indeed, in distinction to the cuprates, broken time-reversal symmetry
below $T_{c}$ was observed in muon spin relaxation \cite{Luke1998,Grinenko2020}
and polar Kerr effect \cite{Xia2006} measurements, reminiscent of
the $p_{x} + ip_{y}$ pairing state of the A-phase of $^{3}$He.

Recently, several experiments have forced the community to reexamine widely accepted beliefs about this fascinating material. NMR measurements revealed  that $\mathrm{Sr}_{2}\mathrm{Ru}\mathrm{O}_{4}$ is in fact a singlet superconductor \cite{Pustogow2019,Ishida2020,Chronister2020}. Applying uniaxial strain leads to a separate onset of superconductivity at $T_c$ and of time-reversal symmetry breaking (TRSB)  at $T_{\TRSB}$ \cite{Grinenko2020}. Combined with the observed jump of certain elastic constants across $T_c$ \cite{Ghosh2020,Benhabib2020}, these observations have led to the proposal that $\mathrm{Sr}_{2}\mathrm{Ru}\mathrm{O}_{4}$ is a two-component singlet superconductor, with $T_{\TRSB}$ signalling the condensation of the second component \cite{Mazin2005, Romer2019,Suh2020,Kivelson2020,Willa2020,Scaffidi2020}. Meanwhile, there is no indication of a heat capacity anomaly at $T_{\TRSB}$ \cite{Li2020}, in contrast to the sharp jump seen at $T_c$. More generally, there is no experimental evidence that time-reversal symmetry breaking is a bulk effect---in fact, muon spin rotation ($\mu$SR) experiments find variations of the effect for distinct samples  and  signatures consistent with diluted magnetic moments \cite{Grinenko2020}. This calls into question the widely accepted view that $\mathrm{Sr}_{2}\mathrm{Ru}\mathrm{O}_{4}$ is a multi-component superconductor.

\begin{figure}
\includegraphics[scale=0.5]{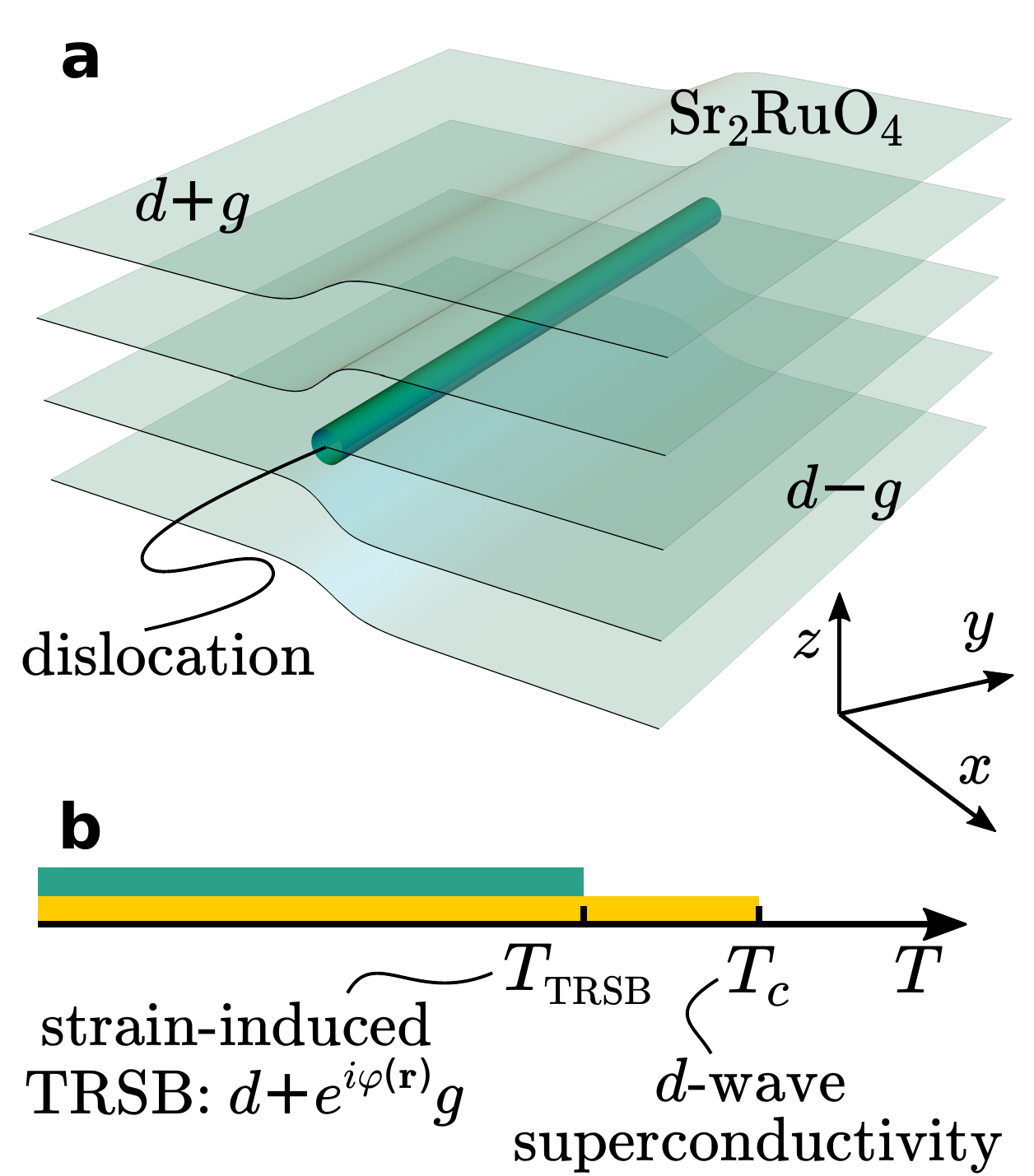}
\caption{
\textbf{(a)}
Edge dislocation in $\mathrm{Sr}_{2}\mathrm{Ru}\mathrm{O}_{4}$ inducing local breaking of time-reversal symmetry due to inhomogeneous strain. Near the dislocation, large strain fields mix a primary $d_{x^{2}-y^{2}}$ pairing
state with other sub-leading symmetry channels, e.g. a $g_{xy(x^{2}-y^{2})}$ pairing
state \cite{Kivelson2020}. As a result, $d \pm g$ pairing emerges on opposite sides of the dislocation, forcing the relative phase $\varphi(\boldsymbol{r})$ to wind in between, thus breaking time-reversal symmetry locally.
\textbf{(b)}
Time-reversal symmetry breaking appears in the bulk system as a stand-alone phase transition where the local phase windings are locked through the long-range decay of strain fields that couple dislocations.}
\label{fig:sketch}
\end{figure}

In this paper, we argue that the pairing states of $\mathrm{Sr}_{2}\mathrm{Ru}\mathrm{O}_{4}$  and the cuprate superconductors display significantly more parallels than previously thought. 
Specifically, we argue that $\mathrm{Sr}_{2}\mathrm{Ru}\mathrm{O}_{4}$
without strain inhomogeneities is a single-component, time-reversal
symmetric superconductor with a $d_{x^{2}-y^{2}}$ pairing
state. This is consistent with recent quasiparticle interference measurements \cite{Sharma2020}
that strongly favor such a pairing state, despite the onset of TRSB. The recently discussed phase diagram of strained $\mathrm{Sr}_{2}\mathrm{Ru}\mathrm{O}_{4}$, where superconductivity occurs in the vicinity of a state with magnetic order, likely a spin-density wave, \cite{Grinenko2020} is also suspiciously similar to the generic cuprate phase diagram with a state of antiferromagnetic order vanishing near superconductivity.

In our theory, time-reversal symmetry breaking of $\mathrm{Sr}_{2}\mathrm{Ru}\mathrm{O}_{4}$ is a consequence of strong strain inhomogeneities that locally break the crystalline symmetry and, due to the large mechanical stresses, strongly couple to a sub-leading pairing state. We analyze the nucleation of the corresponding pairing wave-function and its non-trivial phase windings near edge dislocations of the crystal. In particular, we solve for the pairing state near a single edge dislocation and proceed with a coarse-grained model of many such dislocations. The first part of the theory demonstrates that inhomogeneous strain induces local phase windings at a temperature below $T_{c}$, see Fig.~\ref{fig:sketch}(a), the second shows the emergence of global phase locking. This yields a sharp phase transition where time-reversal symmetry is globally broken, see Fig.~\ref{fig:sketch}(b), albeit with a weak heat-capacity anomaly. It hence explains the absence of a calorimetric signature at $T_{\TRSB}$ in strained samples \cite{Li2020}, which is strong evidence against any scenario where breaking of time-reversal symmetry is associated with the opening of a quasiparticle pairing gap.

Despite the similar pairing state, an important difference between $\mathrm{Sr}_{2}\mathrm{Ru}\mathrm{O}_{4}$ and the cuprates is that at least one other pairing state must be sufficiently close in energy to contribute to the
condensate near strong local strain inhomogeneities. The idea that a second competing superconducting instability is likely present in this material was proposed before both on phenomenological and microscopic grounds \cite{Zhang2018, Romer2019, Kivelson2020, Willa2020,Scaffidi2020}.
In distinction to these investigations,  the sub-leading instability of our theory does not need to be fine-tuned to have approximately the same $T_c$ value as that of the leading $d$-wave instability. Moreover, because the dislocation induces strain in all symmetry channels, TRSB occurs independently of the sub-leading state's symmetry---$s$-wave or $g$-wave.
Finally, in our scenario, TRSB does not correspond to the homogeneous condensation of the second pairing state, but to a collective phenomenon in which the phase windings of the sub-leading pairing wave-function induced near a dislocation phase-lock globally. A direct consequence of the assumption of competing superconducting instabilities is that, in a clean sample, the sub-leading pairing state should lead to a Bardasis-Schrieffer excitation \cite{Bardasis1961} in the electronic Raman spectrum. Finally, strain engineering of $\mathrm{Sr}_{2}\mathrm{Ru}\mathrm{O}_{4}$, e.g. via plastically deformation of the sample, which is known to impact dislocations in other perovskites \cite{Hameed2020}, may be used to create and manipulate time-reversal symmetry broken superconducting states.

\section{Results}
\subsection{Overview of the experimental constraints on the pairing state}
\label{sec:exp}

Before we present our theory we summarize some of the relevant experimental
observations and their puzzling implications that motivated our proposal.
Excluding accidental degeneracies, homogeneous TRSB at $T_{c}$ may only occur for a superconducting
order parameter that has at least two components, i.e., transforms
according to a two- or higher-dimensional irreducible representation
of the symmetry group \cite{Sigrist1991}. For $\mathrm{Sr}_{2}\mathrm{Ru}\mathrm{O}_{4}$, with $D_{4h}$ point group,
this leaves only two options: $E_{u}$ triplet
pairing that transforms like $p_{x} \pm ip_{y}$ or the $E_{g}$ singlet
pairing of the type $d_{xz} \pm id_{yz}$. The most recent NMR measurements
strongly favor singlet pairing \cite{Pustogow2019,Ishida2020,Chronister2020}. Furthermore,
$\mu$SR experiments under uniaxial stress report a clear
splitting between the superconducting transition temperature $T_{c}$ and the onset
temperature for broken time-reversal symmetry, $T_{\TRSB} \!<\! T_{c}$ \cite{Grinenko2020}.
For such a two-component order parameter, uniaxial stress should indeed lift the degeneracy.
In this picture, the strain-favored component first
condenses to a single-component state at $T_{c}$, while the strain-disfavored
component mixes-in with relative phase $\pm\pi/2$ at $T_{\TRSB}$.
Numerous theories for $\mathrm{Sr}_{2}\mathrm{Ru}\mathrm{O}_{4}$ consider either
$p_{x} \pm ip_{y}$ triplet pairing \cite{Rice1995, Baskaran1996, Nomura2000,
Wang2013, Raghu2010, Scaffidi2014}
or, given the recent NMR results of Refs.~\cite{Pustogow2019, Ishida2020,Chronister2020},
$d_{xz} \pm id_{yz}$ singlet pairing \cite{Mazin2005,Huang2019, Suh2020}. 

There is however one very robust thermodynamic argument against
either of these mean-field scenarios (see also appendix~\ref{appendixA}):
recent heat-capacity measurements in strained samples \cite{Li2020}
see no jump at $T_{\TRSB}$ and a large discontinuity at $T_{c}$. Within mean-field theory, the ratio
of the heat-capacity jumps at $T_{c}$ and $T_{\TRSB}$ are generally
related to the slopes $\frac{dT_{c}}{d\epsilon_{B_{1g}}}$ and
$\frac{dT_{\TRSB}}{d\epsilon_{B_{1g}}}$, see Ref.~\cite{Grinenko2020}.
If the magnitudes of the slopes are assumed to be approximately equal for small strain,
the heat-capacity jumps must essentially be the same. In the more
realistic case, as determined by the experimental strain-temperature phase diagram, where $\big|\frac{dT_{\TRSB}}{d\epsilon_{B_{1g}}}\big|<\frac{dT_{c}}{d\epsilon_{B_{1g}}}$,
the jump at the lower transition must even be the largest one. This
is in sharp contrast to the measurements of Ref.~\cite{Li2020}, which reports no anomaly at the lower transition temperature, estimating an upper bound for the second discontinuity at $T_{\TRSB}$ 
of less that $5\%$ of the discontinuity $\Delta C(T_{c})$ at $T_{c}$. This finding
provides strong thermodynamic evidence against a BCS-type transition
of either $E_{g}$ or $E_{u}$ symmetry. More generally, the absence
of a heat-capacity jump at $T_{\TRSB}$ argues against any
scenario where the opening of a pairing gap affects the entropy of
electronic quasiparticles. In addition, the quadratic
dependence of $T_{c}$ with strain \cite{Hicks2014, Watson2018} and the
absence of a jump in the elastic constant in the $B_{1g}$ symmetry
channel \cite{Ghosh2020} constitute additional evidence against a
two-component order parameter with $E_{g}$ or $E_{u}$ symmetry.
For further discussions about the puzzles related to the chiral $E_{g}$
or $E_{u}$ pairing states, see Refs.~\cite{Mineev2007,Kallin2012,Mackenzie2017,Kivelson2020}.

Given this strong evidence against a symmetry-protected, two-component
order parameter, the notion of an accidental (near)-degeneracy of two single-component
order parameters seems a very compelling approach \cite{Zhang2018, Romer2019, Kivelson2020, Willa2020,Scaffidi2020}.
This is rooted in significant microscopic evidence that electronic
pairing mechanisms give rise to several competing superconducting
states \cite{Schnell2006, Cho2013, Zhang2018, Romer2019}.
A range of arguments is brought forward in Ref.~\cite{Kivelson2020} supporting the degeneracy between $\psi_{d}=d_{x^{2}-y^{2}}$
and $\psi_{g}=g_{xy(x^{2}-y^{2})}$ pairing. This state is further
supported by the observed jump of the elastic constant
in the $B_{2g}$ symmetry channel \cite{Ghosh2020, Benhabib2020}, see also Ref.~\cite{Okuda2002}, and by the report of vertical line nodes \cite{Hassinger2017}.
Other candidates include the near degeneracy between
$\psi_{d}=d_{x^{2}-y^{2}}$ and a nodal $s$-wave state $\psi_{s}$ \cite{Romer2019} or the degeneracy between singlet and triplet states in a model that focuses on the quasi one-dimensional parts of the electronic structure \cite{Scaffidi2020}.
Notice, an implication of the small heat-capacity signature at the lower transition is that the amplitude of this second component must be exceedingly small; see appendix~\ref{appendixA}

That the primary pairing state is indeed unconventional is most plainly
demonstrated by the system's strong sensitivity to non-magnetic disorder
 \cite{Mackenzie2003, Mackenzie1998}.
It is further supported by thermal conductivity measurements~\cite{Hassinger2017}, which report vertical nodal lines, and by the recent quasiparticle
interference measurements of Ref.~\cite{Sharma2020}.
It is also in line with numerous microscopic theories for unconventional
pairing in this compound \cite{Schnell2006, Cho2013, Gingras2019, Roising2019, Sharma2020, Scaffidi2020, Husain2020}.
Our approach---partially inspired by Ref.~\cite{Kivelson2020}---relies on the existence of a secondary order parameter that is  reasonably close in energy and that admixes to the primary phase in the presence
of inhomogeneous strain. However, in distinction to Ref.~\cite{Kivelson2020}, we
neither require an almost perfect
degeneracy of both states for unstrained homogeneous samples,  nor
do we expect a second homogeneous mean-field like transition anywhere below the
onset of superconductivity at $T_{c}$. Instead, we argue that the
breaking of time-reversal symmetry is exclusively a consequence of strain inhomogeneities in the
material.

The presence of inhomogeneous strain fields was clearly established
in scanning SQUID microscopy measurements on $\mathrm{Sr}_{2}\mathrm{Ru}\mathrm{O}_{4}$ single crystals \cite{Watson2018}.
The strong coupling of edge dislocations to superconductivity was
revealed in Ref.~\cite{Ying2013}, where the local superconducting
transition temperature in the vicinity of a lattice dislocation can
reach almost twice the value of the bulk. Finally, we note that there
is really no evidence at this point that the TRSB transition in $\mathrm{Sr}_{2}\mathrm{Ru}\mathrm{O}_{4}$
is a mean-field homogeneous phase transition of the bulk system. The $\mu$SR signal shows a broad distribution of local magnetic
moments, somewhat similar to what is often observed in spin glasses \cite{Luke1998,Grinenko2020}. In fact, the authors
of Ref.~\cite{Grinenko2020} explicitly state that the internal field is thought
to arise at edges, defects, and domain walls. In addition, the splitting
of $T_{c}-T_{\TRSB}$ determined in Ref.~\cite{Grinenko2020}
varies in magnitude \emph{and} sign between different samples. These
observations guided us to investigate the role of inhomogeneous strain
fields. We will first analyze the behavior in the vicinity of a single
edge dislocation. In a second step we will formulate an effective
model to describe the case of many dislocations.

\subsection{Time-reversal symmetry breaking at a single dislocation}
\label{sec:single-dislocation}
\begin{figure*}[t]
\includegraphics[width=.95\textwidth]{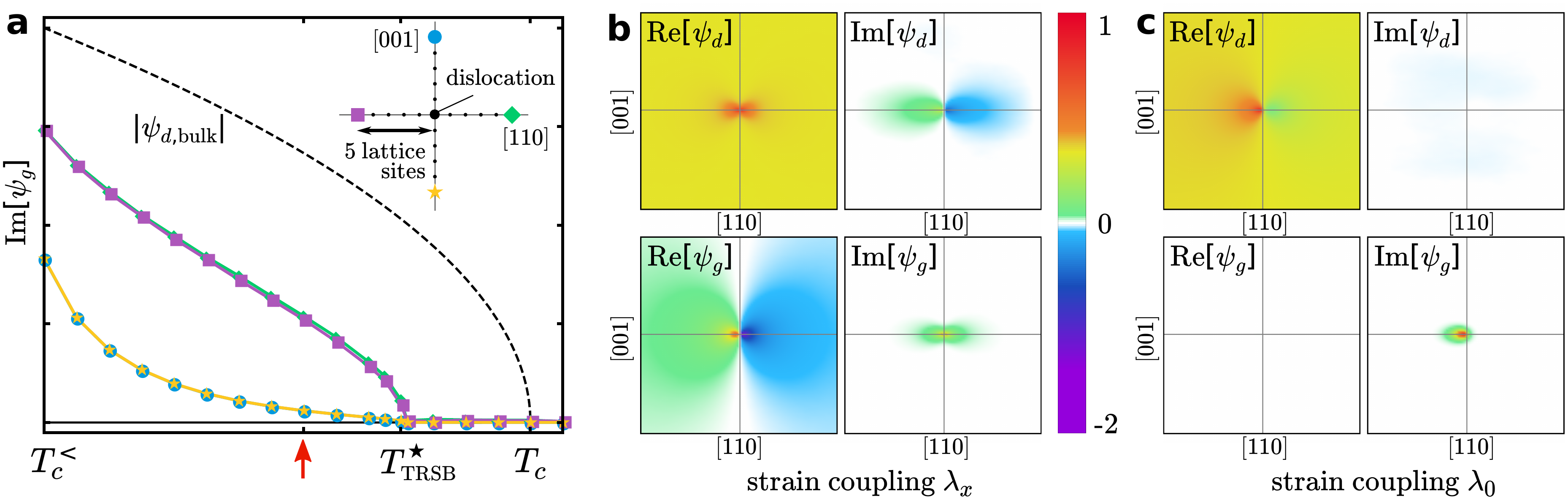}
\caption{Monte-Carlo simulation to minimize the Ginzburg-Landau free energy,
Eq.~\eqref{eq:total-f}, for the two order parameters in the presence of a dislocation.
\textbf{(a)}
Strain-induced onset of the imaginary part of the order parameter $\psi_{g}$ at four locations
(see inset) around a dislocation. Because $\psi_{d}$ is already present at $T < T_c$,
this onset \emph{locally} breaks time-reversal symmetry at $T_{\TRSB}^{\star}$. Note that both these temperatures are much larger than the temperature $T_c^<$ at which 
the homogeneous $g$-wave order parameter would onset.
\textbf{(b)-(c)} Order parameter distribution near the dislocation for
fixed temperature [red arrow in (a)] and for $\lambda_{0} = 5$ and $\lambda_{x} = 8$ respectively, see Eq.~\eqref{eq:strain-coupling}.}
\label{fig:localTRSB}
\end{figure*}
In what follows, we consider a Ginzburg-Landau theory for the superconducting
order parameter $\psi=(\psi_{d},\psi_{g})$, where the
two components belong to different one-dimensional irreducible representations,
i.e.\ they do not transform into each other under a crystalline symmetry.
As discussed, we have in mind a primary order in the $d_{x^{2}-y^{2}}$
channel. For a homogeneous system this order parameter alone would be sufficient; the secondary component only becomes important once we include strain inhomogeneities.  Following Ref.~\cite{Kivelson2020}, this second component is assumed
to be $g_{xy(x^{2}-y^{2})}$. We later discuss how our results change for different sub-leading pairing states, such as $s$-wave.

In an otherwise clean system, such as superconducting $\mathrm{Sr}_{2}\mathrm{Ru}\mathrm{O}_{4}$,
inhomogeneous strain is most prominently induced by dislocations,
i.e. a linear crystallographic defect, as shown in Fig.~\ref{fig:sketch}. To study the impact of dislocation-induced strain on superconductivity we consider the
Ginzburg-Landau free-energy density 
\begin{align}
f=f_{0}+f_{\mathrm{disl}}\label{eq:total-f},
\end{align}
where the dislocation-free contribution $f_{0}$ reads
\begin{align}
f_{0} & = \frac{1}{2}\psi^{\dagger}(r_{0}\tau_{0}-\delta T\,\tau_{z})\psi+\frac{u_{+}}{8}(\psi^{\dagger}\tau_{0}\psi)^{2}+\frac{u_{-}}{8}(\psi^{\dagger}\tau_{z}\psi)^{2}\nonumber \\
 &\quad - \frac{v}{8}(\psi^{\dagger}\tau_{x}\psi)^{2}+\frac{w}{4}(\psi^{\dagger}\tau_{0}\psi)(\psi^{\dagger}\tau_{z}\psi) + f_{\mathrm{grad}}.\label{eq:GL-hom}
\end{align}
Here, $r_0 = T  - T_0$ is the effective temperature, with $T_0$ setting a transition temperature scale. The parameters $u_{\pm}$, $v$, and $w$ are associated with the different symmetry-allowed quartic terms. The conditions $0 < w^{2} < u_{+}u_{-}$
ensure the stability of the functional. The Pauli matrices $\tau_{i}$ act in the space of the two-component order parameter $\psi$. Within a mean-field approach, the system described by Eq.~\eqref{eq:GL-hom} develops a uniform $d$-wave state below $T_{c}\equiv T_{0}+\delta T$,
where it picks a global $U(1)$ phase $\Phi$.
Owing to the invariance of the free energy under the transformation $\psi_{d} \to \psi_{d}$ and $\psi_{g} \to \psi_{g}^{*}$, the relative phase $\varphi$ of the gap function
\begin{align}\label{}
   \Psi = e^{i\Phi} ( |\psi_{d}| + e^{i \varphi} |\psi_{g}| ),
   \label{eq:rel_phase}
\end{align}
becomes a $\mathbb{Z}_{2}$ Ising degree of freedom that can undergo its own transition. In fact, a finite expectation value  of the pseudospin
\begin{align}
   \sigma = \mathrm{sign}( \sin\varphi),
\label{Isingvariable}
\end{align}
marks the onset of the time-reversal symmetry broken state. The condensation of the corresponding degree of freedom plays an important role in our subsequent analysis, albeit not for a homogeneous order parameter.

A uniform $g$-wave component $\psi_{g}$ would emerge below $T_{c}^{<}\equiv T_{0}-\delta T\frac{u_{+}+w}{u_{-}+w}$, with, for $v>0$, a phase $\varphi = \pm \pi/2$ relative to $\psi_{d}$.
In the fine-tuned system where $\delta T=0$, analyzed
in Ref.~\cite{Kivelson2020}, both order parameters develop at the same
temperature.
In this work we consider $\delta T \sim T_c$ sufficiently large such
that the homogeneous, perfect crystal only displays one phase transition
with order parameter $\psi_{d}$.
Still, the expansion Eq.~\eqref{eq:GL-hom} implicitly assumes that the sub-leading
channel $\psi_{g}$ is much closer to condensation than any other pairing channel.
We include the gradient terms 
\begin{align}
\label{gradient}
f_{\mathrm{grad}} = \frac{\kappa_{\parallel}}{2}\big(|\partial_{x}\psi|^{2}+|\partial_{y}\psi|^{2}\big)+\frac{\kappa_{\perp}}{2}|\partial_{z}\psi|^{2},
\end{align}
where $\kappa_{\parallel}>\kappa_{\perp}$ reflects the electronic anisotropy of
the material. In principle, there are additional symmetry-allowed
gradient terms that mix the two components but that will not qualitatively
alter the findings of our analysis.

We now analyze the effect of a single edge dislocation on the pairing
state described by Eq.~\eqref{eq:GL-hom}. Edge dislocations occur even
in otherwise exceptionally clean materials and lead to strong local
strain fields, yielding associated stress values of several GPa \cite{Hull1984}. These
strain fields change the local energetics of the superconductor and
locally break the tetragonal lattice symmetry, mixing different pairing channels.
The strain field in the vicinity of an edge dislocation is given in
standard textbooks \cite{Landau1986}. It is characterized by the Burgers
vector $\boldsymbol{b} = b\, \boldsymbol{\hat{b}}$ and the unit vector $\boldsymbol{\hat{t}}$ tangential
to the dislocation line. The strain tensor $\epsilon_{\alpha\beta} \!=\! (\partial_{\alpha}u_{b}+\partial_{\beta}u_{\alpha})/2$ is evaluated from the displacement field $\boldsymbol{u}$. For an edge dislocation, the latter is given by 
\begin{align}
\!
\boldsymbol{u} &= \frac{\boldsymbol{b}}{2\pi}\Big[\theta \!+\! \frac{A}{2}\sin(2\theta)\Big]+\frac{\boldsymbol{\hat{t}}\times\boldsymbol{b}}{2\pi}[A\cos^{2}(\theta) \!+\! B\log(\rho)]\label{eq:displacement}
\end{align}
where $\rho$ and $\theta$ are the polar coordinates in the plane
perpendicular to the dislocation and the parameters $A$ and $B$ can be expressed
in terms of elastic constants \cite{Landau1986}. While the electronic
system is very anisotropic, the elastic deformations induced by a
dislocation are expected to be similar to those in an isotropic elastic system. The key implications of
Eq.~\eqref{eq:displacement} for the strain tensor $\epsilon_{\alpha\beta}$ are that its magnitude
decays like $1/\rho$, while the sign of the tensor elements depends
on the orientation of $\boldsymbol{b}$ and $\boldsymbol{\hat{t}}$, similarly to the case of an electric field generated by a dipole.
Quite generally then, one finds tension ($\epsilon_{\alpha\beta}>0$) on one side of the dislocation and compression ($\epsilon_{\alpha\beta}<0$) on the
other side, following a typical dipolar angular dependence, see Ref.~\cite{Landau1986}. Based on the results
of Ref.~\cite{Ying2013}, we consider a Burgers vector $\boldsymbol{b}=b \boldsymbol{\hat{z}}$,
see Fig.~\ref{fig:sketch}. Dislocations form loops or end at the sample surfaces. In what follows we consider a  segment that can be considered locally as a straight dislocation line. In our specific analysis,
we opt for a tangent $\boldsymbol{\hat{t}}=(\boldsymbol{\hat{x}}+\boldsymbol{\hat{y}}) / \sqrt{2}$
along the crystallographic $[110]$ direction.
Except for $\boldsymbol{\hat{t}}=\boldsymbol{\hat{x}}$ (or $\boldsymbol{\hat{y}}$)---in which case $\epsilon_{xy}$ vanishes---any generic direction of $\boldsymbol{\hat{t}}$ yields a similar behavior

\begin{figure}[tb]
\includegraphics[width=0.47\textwidth]{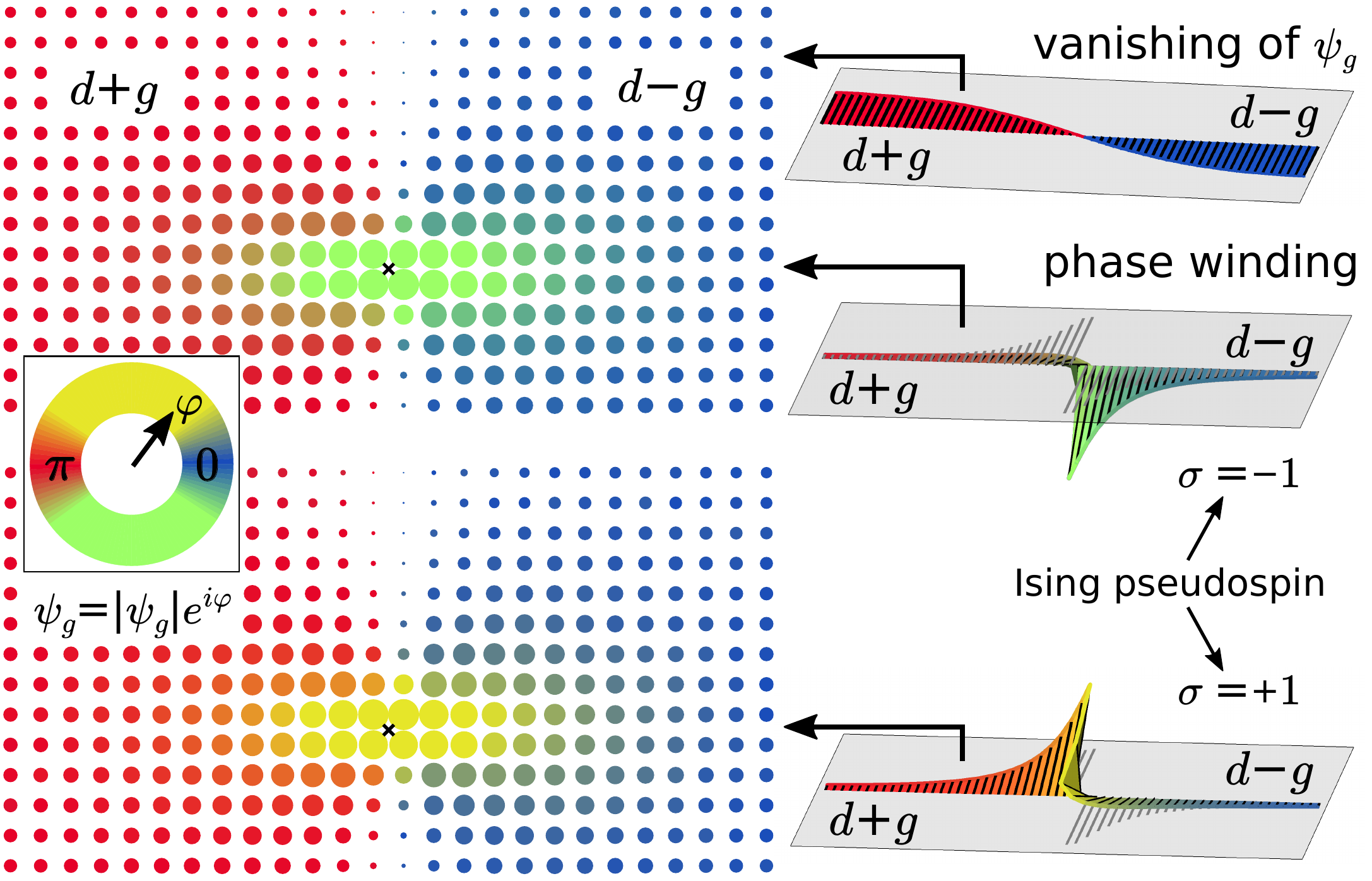}
\caption{
Amplitude (circle size) and phase (color) of the strain-induced $g$-wave component in proximity to a dislocation. Here we consider $\lambda_{x}$ to be the dominant strain-superconductivity coupling, as shown in Fig.~\ref{fig:localTRSB}(b). Inhomogeneous TRSB occurs when the phase winding (right, middle and bottom) is energetically favored over the vanishing of the $g$-wave component. Away from the dislocation the magnitude of the $g$-component is small and a direct sign-change is always preferred (right, top, cut at $z = 25b$). Both degenerate solutions---distinguished by the pseudospin $\sigma$, see Eq.~\eqref{Isingvariable}, and reflecting the Ising symmetry of the TRSB order---are shown.
}
\label{fig:phase}
\end{figure}

The coupling between the dislocation and the superconducting order parameter can be cast
in the form 
\begin{align}
f_{\mathrm{disl}}=\frac{1}{2}\sum_{\alpha=0,x,z}\lambda_{\alpha}\varepsilon^{\alpha}(\boldsymbol{r})\psi^{\dagger}(\boldsymbol{r})\tau_{\alpha}\psi(\boldsymbol{r}).\label{eq:strain-coupling}
\end{align}
Hereby $\varepsilon^{\alpha}(\boldsymbol{r})$ denotes the
projection of the spatial variation of the strain field to distinct
symmetry sectors ($\alpha=0,x,z$) and $\lambda_{\alpha}$ are the associated coupling
constants.
The $A_{1g}$ strain components $\varepsilon^{0}$ and $\varepsilon^{z}$ are linear combinations of $\epsilon_{xx}+\epsilon_{yy}$ and $\epsilon_{zz}$. As such, they locally change the superconducting transition temperature and the degree of degeneracy between the $d$- and $g$-wave components.
The experiments of Ref.~\cite{Ying2013} demonstrated that a large coupling of this kind is present in SrRu$_2$O$_4$.
In our case, the component $\varepsilon^{x}$ corresponds to a $B_{2g}$ strain, $\varepsilon^{x} \equiv \epsilon_{xy}$, which breaks the tetragonal lattice symmetry near the dislocation to a degree that both order parameters belong to the same irreducible representation and, hence, emerge simultaneously. The correspondence $\varepsilon^{x} \leftrightarrow \epsilon_{xy}$ is dictated by the product representation $B_{2g} \!=\! B_{1g} \otimes A_{2g}$ of the order parameters, see Eq.~\eqref{eq:strain-coupling}.
Given the low symmetry of the dislocation-induced strain for a generic orientation of $\boldsymbol{\hat{t}}$ one expects that all strain components are finite. 
The pronounced strain dependence of the bulk superconducting transition \cite{Hicks2014,Watson2018,Grinenko2020} and the near doubling of the local onset of superconductivity near dislocations~\cite{Ying2013} are evidence for strong coupling constants $\lambda_\alpha$.
It is instructive to investigate the mechanism of local TRSB for pure strain couplings $\propto \lambda_{\alpha}$. For either $\lambda_{0}$ or $\lambda_{z}$ the breaking of time-reversal symmetry is dictated by the quartic term $-v(\psi^{\dagger}\tau_{x}\psi)^{2}/8$ in Eq.~\eqref{eq:GL-hom}. This implies that the strain-induced $g$-wave component develops with a phase $\pm \pi/2$ relative to the leading order.
In distinction, the coupling $\propto\lambda_{x}$ mixes the two order parameters in a way that on opposite sides of the dislocation we find $d + g$ and $d - g$ pairing states, since the strain changes from tensile to compressive as one traverses the dislocation. To connect the $d + g$ and $d - g$ regions, the $g$-wave component can either vanish or establish a local time-reversal symmetry broken state $d+e^{i\varphi(\boldsymbol{r})}g$ where the phase $\varphi$ gradually winds from $0$ to $\pi$.
The system opts for the second possibility when the strain term $\pm \lambda_{x} \cos\varphi$ dominates over the gradient terms in Eq.~\eqref{gradient}.
The mechanism is reminiscent, albeit qualitatively different, from other cases discussed in the literature where time-reversal symmetry breaking superconductivity takes place at twin boundaries of the crystal \cite{Callin1997,Sigrist1998}.

Independent of the strain coupling, the onset of a non-trivial phase winding takes place at a well-defined temperature $T_{\TRSB}^{\star}$.
Indeed, the problem of determining the onset of phase-winding near a dislocation can be recast as a problem of finding the condition under which a bound state emerges in a Schr\"odinger-type problem \cite{Massih2011};  see appendix~\ref{appendixB}. The onset temperature of local TRSB is then larger than the temperature where a homogenous $\psi_g$ component would appear, $T_{\TRSB}^{\star} \geq T_{c}^{<}$. 
Importantly, $f + f_\mathrm{disl}$ is still invariant under the transformation $\psi_{d} \to \psi_{d}$ and $\psi_{g} \to \psi_{g}^{*}$ and hence, the $\mathbb{Z}_{2}$ Ising degree of freedom remains intact.
Thus, at $T_{\TRSB}^{\star}$ the system locally picks one of two degenerate solutions, solely distinguished by the Ising pseudospin $\sigma = \pm 1$, see Eq.~\eqref{Isingvariable}.
%
A large system with many dislocations undergoes a TRSB transition at $T_{\TRSB}$ once it locks all its Ising pseudospins $\sigma_{i}$ in a ferromagnetically ordered state and \emph{globally} breaks the $\mathbb{Z}_{2}$ symmetry.
This point will be further explored in the next section, where we will show that $T_{\TRSB} \lesssim T_{\TRSB}^{\star}$.

We have solved the coupled non-linear equations $\delta f / \delta\psi_{d}=\delta f / \delta\psi_{g}=0$
for the total free energy Eq.~\eqref{eq:total-f} via a stochastic annealing approach. All numerical results are obtained from a Monte-Carlo simulation of a
discretized system with $100\times100$ lattice sites (in units of $b$) with
$(u_{+} \!+\! u_{-})/2 \!=\! v \!=\! 1$,
$(u_{-} \!-\! u_{+}) / 2 \!=\! w \!=\! 0.5$, $\kappa_{\parallel} \!=\! 10$,
$\kappa_{\perp} \!=\! 2$.
A realistic description of the strain fields is achieved with the simple expression $\varepsilon^{\alpha}(\boldsymbol{r})=\varepsilon^{\alpha}(\rho,\theta)=b\cos(\theta)/\rho$.
The spatial distribution of the order parameters near
the dislocation is shown in Fig.~\ref{fig:localTRSB} for the limiting cases where the strain-superconductivity coupling
is purely determined by $\lambda_{x}$ [Fig.~\ref{fig:localTRSB}(a,b)] and by $\lambda_{0}$ [Fig.~\ref{fig:localTRSB}(c)], respectively.
Fig.~\ref{fig:phase} illustrates both options of connecting the $d \pm g$ regions: near (far from) the dislocation the connection occurs via phase winding (vanishing) of $\psi_{g}$. Furthermore, it also shows the two degenerate solutions $\sigma = \pm 1$ for winding the phase between the $d + g$ and $d - g$ regions.

\begin{figure}[tb]
\includegraphics[width=0.45\textwidth]{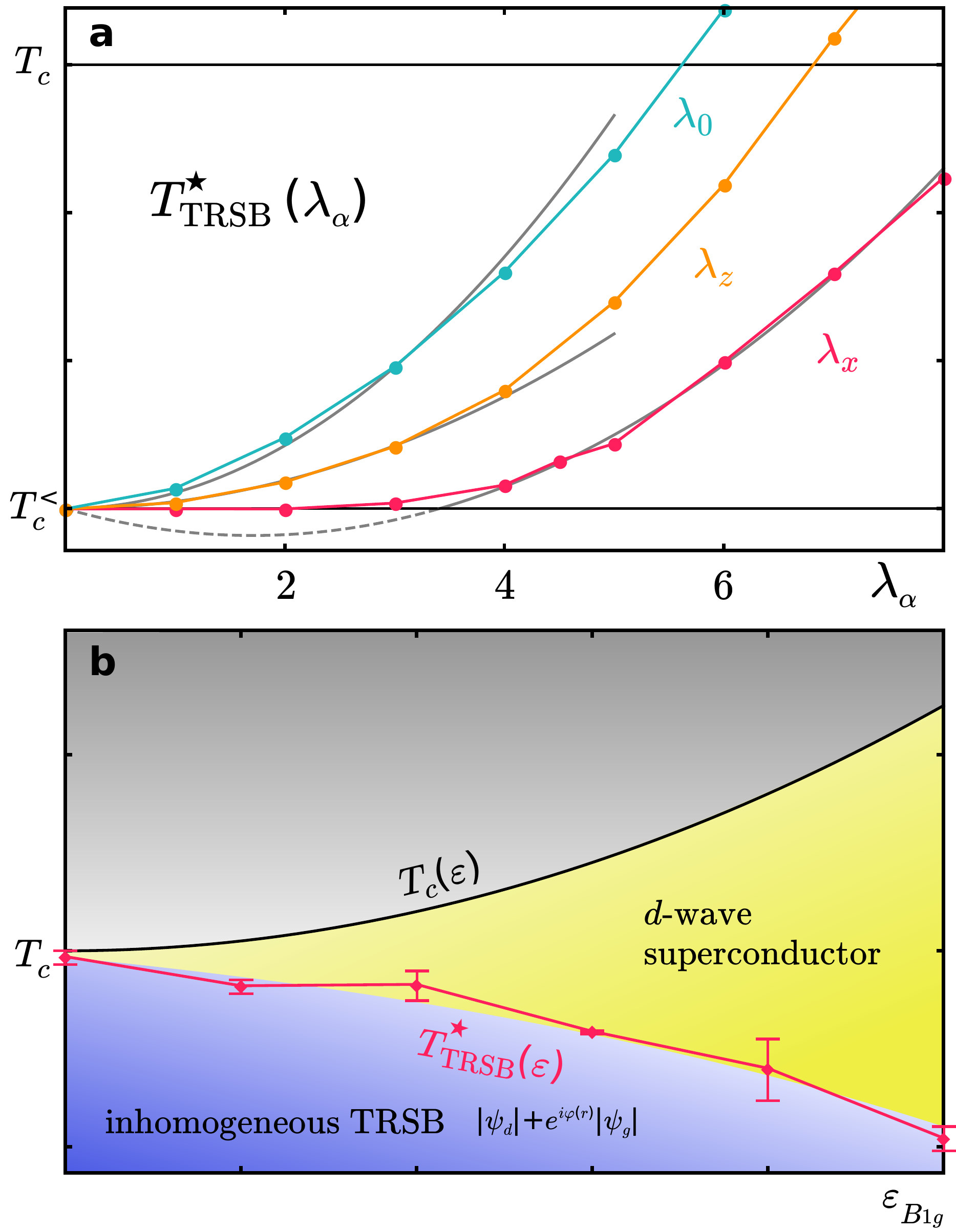}
\caption{\textbf{(a)} Evolution of $T_{\TRSB}^{\star}$ upon varying the interaction parameters $\lambda_{0}$, $\lambda_{z}$, and $\lambda_{x}$. Recall that $T_c^<$ is the temperature at which a homogeneous $\psi_d$ component would onset, resulting in a homogeneous time-reversal symmetry broken state. The analytic results (grey curve) indicate that $T_{\TRSB}^{\star}(\lambda_{0,z})$ deviates quadratically away from $T_{c}^{<}$. Meanwhile, $T_{\TRSB}^{\star}(\lambda_{x})$ deviates from $T_c^<$ only above a threshold value.
\textbf{(b)} Phase diagram as function of homogeneous $B_{1g}$ strain and temperature. It is evaluated for $\lambda_{x}=8.75$ and for $T_{c}(0) = T_{\TRSB}^{\star}(0)$. The analytic relation $T_{c}(\epsilon) - T_{c}(0) = (\delta T^{2}+\epsilon^{2})^{1/2}$ is shown in black.}
\label{fig:lambda-epsilon-dependence}
\end{figure}

The dependence of $T_{\TRSB}^{\star}$ on the magnitude
of the coupling constants $\lambda_{x}$ and $\lambda_{0}$ is shown
in Fig.~\ref{fig:lambda-epsilon-dependence}(a). For sufficiently large strain couplings one
can induce local TRSB pairing above the bulk temperature, a
consequence of the fact that pairing can be boosted locally. 
In Fig.~\ref{fig:lambda-epsilon-dependence}(b) we analyze
the change of the bulk transition temperature $T_{c}$ and of the local
TRSB onset temperature $T_{\TRSB}^{\star}$ as a function
of the in-plane strain $\epsilon_{B_{1g}}$.  Fig.~\ref{fig:lambda-epsilon-dependence} was obtained for the
specific coupling strength $\lambda_{x} \!=\! 8.75$ after fine-tuning $T_{c}
\approx T_{\TRSB}^{\star}$ at zero applied strain. The superconducting transition
temperature changes quadratically with strain,
as is required for a single-component order parameter. $T_{\TRSB}^{\star}$ on the other hand is weakly affected by strain
values that are much smaller than the local, dislocation-induced strain.
Even the largest applied strain field in modern experiments amounts
to stress values that are an order of magnitude smaller than typical
edge-dislocation stresses \cite{Hull1984}. This may explain why the experiments in Ref.~\cite{Grinenko2020} did not observe a significant change for the TRSB onset temperature with applied strain along the $[100]$ direction.

It is also interesting to discuss what happens if strain is applied along the $[110]$ direction (i.e.\ $B_{2g}$ strain). For our scenario of inhomogeneous local TRSB at $T_{\TRSB}^{\star}$, the phase diagram of Fig.~\ref{fig:lambda-epsilon-dependence}(b) would remain essentially unchanged. However, for the scenario of homogeneous TRSB taking place at $T_c^<$, significant changes are expected. In particular, because $B_{2g}$ strain couples the $\psi_d$ and $\psi_g$ order parameters bi-linearly, one would generally expect a stronger, linear enhancement of $T_c$ and of the separation between $T_c$ and the TRSB transition temperature with increasing $B_{2g}$ strain.  

To discuss what happens in the presence of an external magnetic field $H_{z}$ along the $z$-direction, we note that it transforms according to the $A_{2g}$ irreducible representation and that it is odd under time reversal. Consequently, there is an additional term in $f_{\mathrm{disl}}$
\begin{align}
f_{\mathrm{disl}} \rightarrow  f_{\mathrm{disl}} - H_{z} \lambda_{y} \varepsilon^{y}(\boldsymbol{r}) \psi^{\dagger}(\boldsymbol{r}) \tau_{y} \psi(\boldsymbol{r})/2,
\end{align}
where $\varepsilon^{y}=\epsilon_{x^{2}-y^{2}}$ is the projection of the strain field onto the $B_{1g}$ channel. This coupling is crucial to explain the finite Kerr signal observed in Ref.~\cite{Xia2006}, as it allows one to train different regions of broken time-reversal symmetry with a magnetic field. The condition of a broken  vertical mirror symmetry to achieve a finite Kerr signal is naturally fulfilled near the dislocation.  

Let us finally discuss what would happen for different combinations
of single-component order parameters. Consider the case $\psi=(\psi_{d}, \psi_{s})$
with a second component transforming as $A_{1g}$, i.e. $s$-wave. Then, the symmetry-mixing
strain in Eq.~\eqref{eq:strain-coupling} is given by $\varepsilon^{x} = \epsilon_{x^{2}-y^{2}}$
since $B_{1g} \!=\! B_{1g}\otimes A_{1g}$. Other than that, our analysis would
be largely unchanged. On the other hand if one considered the case of 
$d_{x^{2}-y^{2}}$ and $d_{xy}$ states, no strain mixing
term $\varepsilon^{x}$ is allowed, because $A_{2g} \!=\! B_{1g} \otimes B_{2g}$ and there is no $A_{2g}$ strain in the $D_{4h}$ point group.
The same is true for a combination of $s$-wave with $g$-wave pairing.
Hence, the presented scenario applies to combinations of $d$ with
either $g$- or $s$-wave states.
 The time-reversal symmetry-broken state of all combinations of single-component
order parameters can be trained with a magnetic field $H_{z}$, since the strain fields induced by the dislocation always contain a component that couples $H_z$ with the two components of $\psi$.

Thus far, we established the occurrence of phase winding near a single dislocation and determined the onset temperature scale $T_{\TRSB}^{\star}$ for this behavior.  
The local breaking of time-reversal symmetry---marked by a non-vanishing $|\psi_{g}|\sin(\varphi)$---takes place in a finite region around the core of the dislocation. In the next section we allow for the Ising variable $\sigma$ to vary in space along the dislocation line and between different edge dislocations.

\subsection{Collective excitations and global phase transition for many dislocations}
\label{sec:many-dislocations}
\begin{figure}
\includegraphics[width=0.4\textwidth]{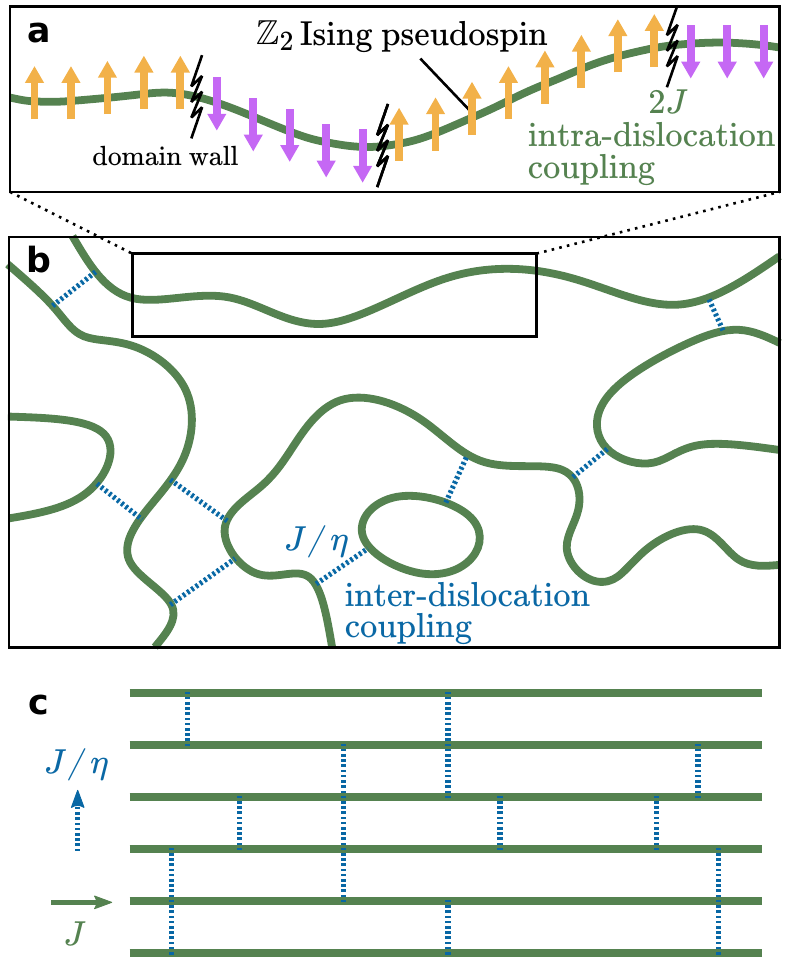}
\caption{
Mapping of the physical system to an effective coarse-grained model. Panel (a) shows the Ising domain walls associated with the Ising TRSB order parameter of Eq. \ref{Isingvariable}, which proliferate along a dislocation line according to the 1D Ising model. Panel (b) illustrates the coupling between multiple dislocations, which can form open lines connecting the surfaces of the sample of closed loops. As a result, as shown in panel (c), the critical behavior of the model is the same as that of an anisotropic 2D Ising model with exchange parameters $J$ and $J/\eta$, with $\eta \gg 1$. }
\label{fig:coarse-grained}
\end{figure}

In this section we analyze the case of many coupled dislocations. We find that this leads to a sharp TRSB phase transition in the Ising universality class at a temperature $T_{\TRSB}$. Given the underlying Ising symmetry, it is however possible that the phase variable undergoes a thermally excited kink $\varphi\rightarrow 2\pi n -\varphi $ along the dislocation line. In terms of the Ising variable $\sigma$, see Eq.~\eqref{Isingvariable}, the kink corresponds to a domain wall, as illustrated in Fig.~\ref{fig:coarse-grained}(a). In a coarse-grained approach, the statistical mechanics of a single dislocation
maps to a one-dimensional Ising chain
\begin{align}\label{}
   H_{1} &= - J \sum_{i}^N \sigma_{i} \sigma_{i+1},
\end{align}
where the dislocation is divided along its length $L$ in $N = L/\ell$ segments of typical length $\ell$. Each segment is associated with a local Ising spin $\sigma_{i}$ and a binding energy per unit length $J$. The characteristic values
\begin{align}
\ell \sim (\kappa_{\parallel} \kappa_{\perp})^{1/2}/\lambda_{\alpha} b,
\label{length}
\end{align}
and $J\sim T_{\TRSB}^{*}$ 
follow from the aforementioned solutions of the Schr\"odinger equations, see appendix \ref{appendixB}. For the one-dimensional Ising model it holds that $\langle \sigma\rangle = 0$ at any finite temperature. The density of kinks per unit length, $n_{\mathrm{kink}} \sim -\ell^{-1} \log[\tanh(2J/T)]$, is exponentially small at low $T$. Shorter dislocation loops, with $N<e^{2J/T}$, have negligible probability for kinks and are essentially ordered. However, averaging over many uncoupled loops gives rise to a vanishing global order parameter $\langle \sigma \rangle$.

To account for the coupling between distinct dislocations, we note that the magnitude of local TRSB decays exponentially as $e^{-\rho/\ell}$ in the direction transverse to the dislocation. Then, two dislocation segments at a distance $\rho$ will interact via the exponentially small coupling $J'(\rho) = J e^{-\rho/\ell}$.
From an analysis of the gradient terms of the Ginzburg-Landau expansion follows that the sign of this interaction is ferromagnetic. The many-dislocation problem is therefore governed by the coarse-grained Hamiltonian
\begin{align}\label{}
   H = \sum_{m} H_{1}^{m}
     + \sum_{m,m'} \sum_{i,j} 
                              J'(\rho_{i,j}) \sigma_{i}^{m}\sigma_{j}^{m'}.
\end{align}
The superscripts $m$, $m'$ label the dislocation, and $\rho_{i,j}$ denotes the distance between the segments $i$ (of dislocation $m$) and $j$ (of dislocation $m'$).

Owing to the exponential dependence of the coupling $J'(\rho)$, one can further map the problem to an anisotropic Ising model with coupling constant $J$ in one direction and a single coupling constant $J/\eta$ in the orthogonal directions. Here $\eta = e^{\bar{\rho}/\ell}$ with $\bar{\rho}$ the typical value for the shortest distance of two neighboring dislocations. This is shown in panels (b) and (c) of  Fig.~\ref{fig:coarse-grained}. The anisotropic Ising model develops a finite order parameter $\langle \sigma \rangle$ below the transition temperature

\begin{align}\label{}
   T_{\TRSB} \sim J / \ln(\eta) \sim (\ell/\bar{\rho})  T_{\TRSB}^{\star}.
\end{align}
This result 
follows regardless of whether the coupling  via $J'$  leads to  a two- or three-dimensional network of dislocations. It
is a consequence of the one-dimensional Ising model being at its lower critical dimension.
The above expression implicitly assumes a weak coupling between nearby dislocations, i.e.\ $\ell < \bar{\rho}$. In that case the transition temperature $T_{\TRSB}$ is smaller than, but potentially comparable to, the local ordering temperature $T_{\TRSB}^{\star}$. While the ordering of single dislocations is hindered by kinks along the loop, the weak dislocation-dislocation coupling locks the global Ising variable and leads to a global symmetry breaking.

\begin{figure}[h!]
\includegraphics[width=0.45\textwidth]{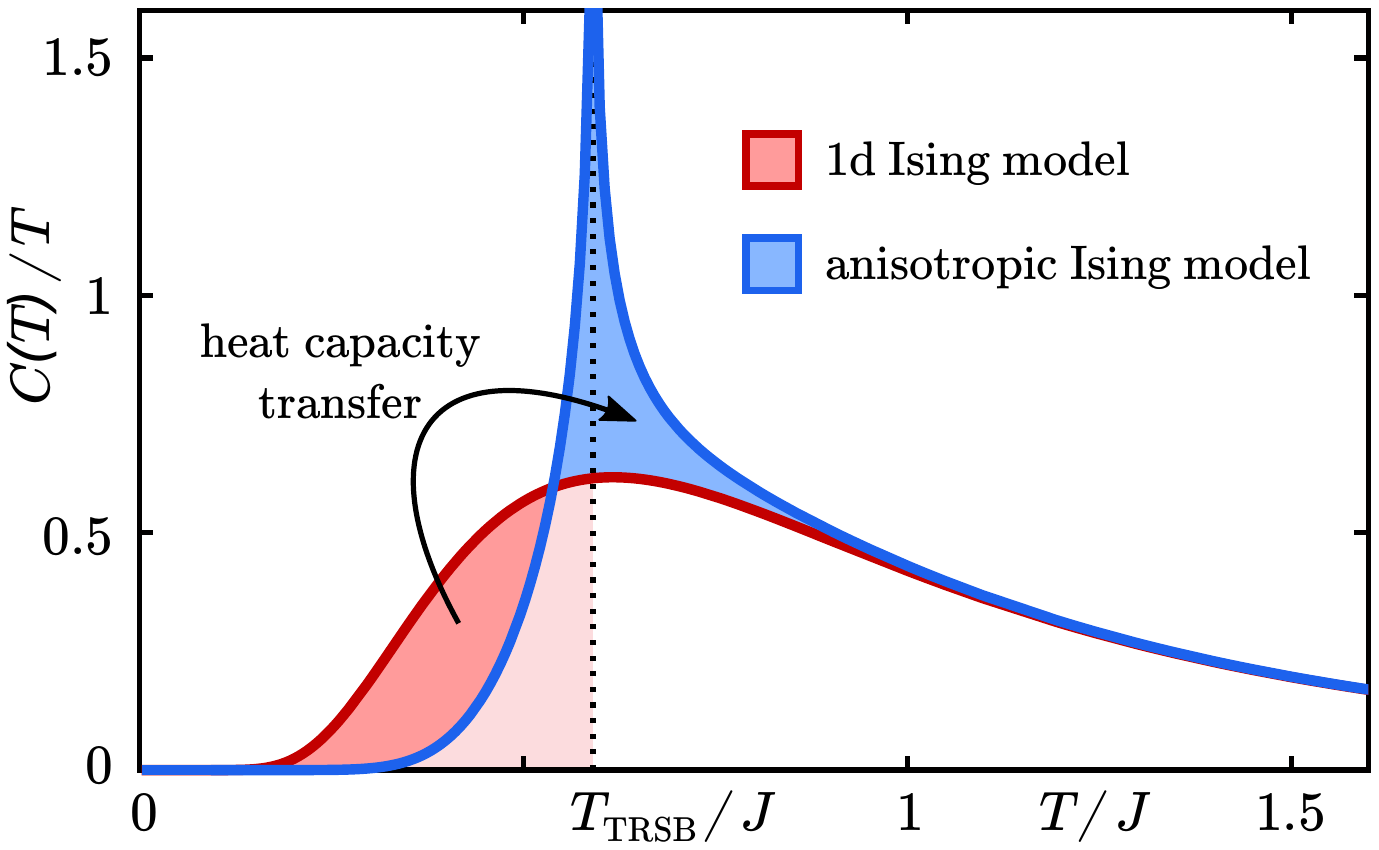}
\caption{
Entropy change $C(T)/T$ per Ising pseudospin (in units of $\kB/J$) of the anisotropic Ising model with $J'/J=0$ (red) and $0.01$ (blue). This illustrates our approach to estimate the total weight of the heat-capacity anomaly at the transition for finite $J'$ from the transferred calorimetric weight in $C_{\mathrm{Ising},d=1}/T$. }
\label{fig:heat_capacity}
\end{figure}

The heat-capacity anomaly associated with the lifting of the $\mathbb{Z}_{2}$ symmetry is much smaller in magnitude than the one due to the opening of the single particle gap at the bulk superconducting transition temperature $T_{c}$.
To show this, we estimate the change of the electronic specific heat at the superconducting transition as $\Delta C(T_{c})\sim S_{\mathrm{el}}(T_{c}) \sim T_c/\eF$, with $S_{\mathrm{el}}$ the electronic contribution to the entropy and $\eF$ the Fermi energy [the Boltzmann constant $\kB$ is set to unity].
In contrast, the heat capacity per Ising spin of an anisotropic two-dimensional Ising model follows from Onsager's solution and is shown in Fig.~\ref{fig:heat_capacity}.
The extent of the calorimetric anomaly at $T_{\rm TRSB}$ [blue shaded area in Fig.~\ref{fig:heat_capacity}] corresponds to the heat-capacity transfer [dark red area in Fig.~\ref{fig:heat_capacity}]. To estimate the magnitude of this effect, we use the approximate form 
\begin{align}\label{eq:approx-deltaS}
  \delta S_{\TRSB}\approx\int_{0}^{T_{\TRSB}}\frac{C_{\mathrm{Ising}, d=1}(T')}{T'}dT',
\end{align}
with the heat capacity of the one-dimensional Ising model
\begin{align}
  C_{\mathrm{Ising}, d=1}=\frac{a^{2}c}{\bar{\rho}^{2}\ell} \frac{J^2}{T^2 [\cosh(J/T)]^{2}}.
\end{align}
The expression in Eq.~\eqref{eq:approx-deltaS} accounts for the total red area in Fig.~\ref{fig:heat_capacity}. Here $a$ and $c$ are the in-plane and out-of-plane lattice constants.
The small prefactor $a^{2}c/(\bar{\rho}^{2}\ell )$ accounts for the number of Ising pseudospins per unit cell and is needed to compare with the electronic contribution. Using Eq.~\eqref{length} and the dimensionless coupling $\bar{\lambda}_{\alpha} = \lambda_{\alpha}/T_{c}$ we find 
$\ell = (a/\bar{\lambda}_{\alpha})(\eF/T_{c})^{2}$ and obtain the ratio
\begin{align}
 \frac{\delta S_{\TRSB}}{S_{\mathrm{el}}(T_{c})} \sim
 \frac{1}{S_{\mathrm{el}}(T_{c})} \frac{a^{2}c}{\bar{\rho}^{2}\ell}\frac{\log\eta}{\eta^{2}}
 \sim
 \bar{\lambda}_{\alpha}^{2} \frac{a}{ \bar{\rho}} \Big(\frac{T_{c} }{ \eF}\Big)^{3} e^{-2\bar{\rho}/\ell}.
\end{align}
As anticipated, the entropy associated with the collective phase locking, and thus to the TRSB transition, is---in addition to the two small prefactors $(T_{c}/\eF)^{3}$ and $(a/\bar{\rho})$---exponentially suppressed.

Our conclusion of a global Ising transition at $T_{\TRSB}$ can also be reached by considering disorder that is caused by dislocations at random positions and with random Burgers-vector orientations~\cite{SAKivelson}. From Eq.~\eqref{eq:strain-coupling} follows that dislocation strain acts like a uniaxial random field that couples to $\cos[\varphi(\boldsymbol{r})]$, with the relative phase $\varphi(\boldsymbol{r})$, see Eq.~\eqref{eq:rel_phase}. The corresponding $XY$-model with random uniaxial field  is a well-studied problem in the statistical mechanics of disordered systems, with several rigorous results~\cite{Aharony1978,Minchau1985,Feldman1998,Feldman1999,Crawford2013,Crawford2014,Bera2014}. It undergoes an Ising transition with order parameter    
$\langle \cos[\varphi(\boldsymbol{r}]) \rangle \neq 0$, fully consistent with our analysis. This demonstrates that our conclusions concerning the nature of the transition as clean Ising transition are robust despite the random nature of the dislocation problem.  

\section{Discussion}
In this paper we proposed an alternative origin for
the observed breaking of time-reversal symmetry in $\mathrm{Sr}_{2}\mathrm{Ru}\mathrm{O}_{4}$ due to inhomogeneous strain
fields that locally couple a single-component primary order parameter
to a sub-leading pairing state. While the material without strain
inhomogeneities would only display the primary order, which is likely
of $d_{x^{2}-y^{2}}$ symmetry, strong strain fields mix these symmetry-distinct
pairing states. A promising candidate for such strain inhomogeneities
are those originated near edge dislocations as shown in Fig.~\ref{fig:sketch}.
Local stress values near such dislocations are easily in the regime
of several GPa, leading to a strong order-parameter mixing. Compressive
and tensile strain on opposite sides of the dislocation are then shown
to lead to local TRSB via phase-winding of the induced pairing wave-function. The main appeal and ultimate motivation for this proposal is the absence of a mean-field behavior
of the heat capacity at $T_{\TRSB}$ under homogeneous strain, while such
behavior occurs at $T_{c}$. 
A scenario
which attributes TRSB to the opening of an additional pairing gap would directly affect the entropy of electronic quasiparticles.
The fact that no heat-capacity jump is experimentally detected at $T_{\TRSB}$ is a strong evidence against any such scenario.

There are certain similarities, but also important differences between
our theory and the description of Ref.~\cite{Kivelson2020} advocating an
accidental degeneracy of homogeneous $d$- and $g$-wave pairing. In
both cases two symmetry distinct states are relevant and time-reversal
symmetry breaking is a consequence of the relative phase between
these two states. In both descriptions one has to fine-tune $T_{\TRSB}\approx T_{c}$
for unstrained samples. There are, however, significant distinctions.
First, in our approach it is in principle possible to observe time-reversal symmetry broken states
via local probes even above $T_{c}$. This effect is a consequence of the
local enhancement of pairing near dislocations, as reported by
experiments \cite{Ying2013}.
Second, our findings suggest that TRSB may be engineered by controlling
the dislocation density during sample preparation, which in principle allows for
a clearer separation between $T_{c}$ and $T_{\TRSB}$.
Evidence for a sample-dependence of $T_{\TRSB}$, and even of $T_{\TRSB}>T_{c}$,
was reported in Ref.~\cite{Grinenko2020}. Third, we do not expect
a strong heat-capacity anomaly at $T_{\TRSB}$. In our description, time-reversal symmetry breaking is a phase locking process where comparatively
few degrees of freedom are involved and no gap-opening of single particle states
occurs. As a result, the transition leaves weak signatures in the heat-capacity signal and hence, naturally explains the absent heat-capacity
anomaly at $T_{\TRSB}$ found in strained samples \cite{Li2020}.
In the case of a homogeneous $d \!+\! ig$ state \cite{Kivelson2020} one would expect a jump in the heat capacity at the second bulk transition. 
Finally, a prediction that may be used to falsify our theory in future
experiments is the change of the transition temperature $T_{c}$ with
uniaxial strain $\epsilon_{B_{1g}}$ and $\epsilon_{B_{2g}}$. We
expect that $T_{c}$ changes quadratically with strain for both symmetries,
while Ref.~\cite{Kivelson2020} clearly predicts a quadratic variation
with $\epsilon_{B_{1g}}$ but a linear change with $\epsilon_{B_{2g}}$.
An investigation of $T_{c}(\epsilon_{B_{2g}})$ should
be able to discriminate between the two scenarios. 

An important observation that concerns the pairing state of $\mathrm{Sr}_{2}\mathrm{Ru}\mathrm{O}_{4}$ is the discontinuity of the elastic constant $C_{66}$ at the superconducting transition temperature \cite{Ghosh2020,Benhabib2020}, see also Ref.~\cite{Okuda2002}. Elastic constants associated with strain that transforms trivially, i.e.\ according to $A_{1g}$, are expected to be discontinuous at any superconducting transition \cite{Testardi1975}. However, elastic constants like $C_{66}$ or $C_{11}-C_{12}$, corresponding to $B_{1g}$ or $B_{2g}$ strain, respectively, are continuous for a single-component order parameter.  A jump of these elastic constants then clearly signals a multi-component order parameter \cite{Fernandes2013}.
Nevertheless, they should not be taken as irrevocable evidence for a homogeneous two-component order parameter. Given the slowly decaying dislocation-induced $g$-wave component, the system is according to our theory essentially a two-component order parameter  for large parts of the sample.  Our analysis of single dislocations shows clearly that the regions of the sample with local $d\pm g$ pairing, that should affect the elastic constant jump, are much larger than the regions near the core of the dislocation, where time reversal symmetry is  broken. This gives rise to a discontinuity of $C_{66}$ while $C_{11}-C_{12}$ should be continuous or at best have a much smaller jump consistent with Refs. \cite{Ghosh2020,Benhabib2020,Okuda2002}. 
Quantitatively, the scattering of elastic waves off dislocations is strong \cite{Li2017_a,Li2017_b} and gives rise to corrections of the elastic constants of the order of $10^{-2}$ \cite{Grimwall1999}. These effects are much larger than the largest observed relative magnitudes of the discontinuities of Ref \cite{Ghosh2020}, which are of the order of $10^{-4}$. 
Finally, the size of the reported jump of $C_{66}$  in different measurements (compare Refs. \cite{Ghosh2020} and \cite{Benhabib2020}) varies significantly, by about two orders of magnitude. While this may  partially be due to the different experimental techniques used, sample-to-sample variations of the effect follow very naturally within our theory.

Our proposed scenario for the origin of TRSB in $\mathrm{Sr}_{2}\mathrm{Ru}\mathrm{O}_{4}$ could also
be tested by experiments in plastically deformed samples. In contrast to elastic strain, plastic strain is known to create, move,
and combine dislocations. It was recently observed that the transition temperature of another superconducting perovskite, $\mathrm{Sr}\mathrm{Ti}\mathrm{O}_{3}$, 
increased substantially upon application of plastic strain, a behavior attributed to
the formation of self-organized dislocation structures \cite{Hameed2020}. In $\mathrm{Sr}_{2}\mathrm{Ru}\mathrm{O}_{4}$, an enhancement of the 
dislocation density by plastic strain would likely result in a larger TRSB signal. Even if the dominant effect of plastic deformation is 
to combine and merge the existing dislocations, one would  still expect a significant change of the TRSB signal. Manipulating this unconventional superconductor via plastic deformation may therefore be utilized to strain engineer the degree of time-reversal symmetry breaking in the sample. 
\vspace{1em}

\begin{acknowledgements}
We are grateful to Erez Berg, Martin Greven, Clifford Hicks, Steven A. Kivelson, Avraham Klein, You-Sheng Li, Andrew P. Mackenzie, Brad Ramshaw, and Andrew Chang Yuan for fruitful discussions. 
R.W., M.H., and J.S. were supported by the Deutsche Forschungsgemeinschaft (DFG, German Research Foundation) - TRR 288-422213477 Elasto-Q-Mat (project B01). R.M.F was supported by the U.S. Department of Energy through the University of Minnesota Center for Quantum Materials, under Grant No. DE-SC-0016371.
\end{acknowledgements}

\appendix

\section{Heat capacity of two degenerate superconductors}
\label{appendixA}
The primary motivation for our analysis was the absence of a signature in the heat capacity at the time-reversal symmetry breaking transition of strained samples \cite{Li2020}, where $\mu$SR experiments find that $T_{c}$ and $T_{\TRSB}$ are distinct transitions \cite{Grinenko2020}. In this appendix we briefly summarize the heat-capacity anomalies for homogeneous two-component order parameters. We use the Ginzburg-Landau expansion of Eq.~\eqref{eq:GL-hom}. For $\delta T=0$ one should distinguish between an order parameter that belongs to a two-dimensional irreducible representation, such as $E_{g}$ or $E_{u}$, and two accidentally degenerate oder parameters that each belong to one-dimensional irreducible representations. From the perspective of the heat capacity, the former is a special case of the latter where the quartic coupling constant $w$ in Eq.~\eqref{eq:GL-hom} vanishes. 

For degenerate order parameters holds $\delta T=0$, and $T_{c}=T_{0}$. The discontinuity of the heat capacity is
\begin{align}
 \frac{\Delta C(T_{0})}{T_{0}}=\frac{u_{-}}{u_{+}u_{-}-w^{2}}.
\end{align}
Let us now analyze finite $\delta T$, i.e. the degeneracy is lifted. The upper transition temperature is $T_{c}=T_{0}+\delta T$ with heat-capacity jump
\begin{align}
 \frac{\Delta C_{1}(T_{c})}{T_{c}}=\frac{1}{u_{+}+u_{-}+2w}.
\end{align}
The lower transition temperature is $T_{c}^{<}=T_{0}-\frac{u_{+}+w}{u_{-}+w}\delta T$ and the corresponding heat capacity jump is 
\begin{align}
 \frac{\Delta C_{2}(T_{c}^{<})}{T_{c}^{<}}=\frac{(u_{-}+w)^{2}}{u_{-}u_{+}-w^{2}}\frac{\Delta C_{1}(T_{c})}{T_{c}}.
\end{align}
Let us first focus on the case when $\delta T$ lifts the degeneracy of a two-dimensional irreducible representation. In this situation it holds $w=0$. It then follows that the ratio of the heat capacities at the two transitions equals the ratio of the slopes of the transition temperatures
\begin{align}
  \frac{\Delta C_{2}(T_{c}^{<})}{\Delta C_{1}(T_{c})} =\frac{u_{-}}{u_{+}}
  = \frac{\frac{dT_{c}}{d\delta T}}{\left|\frac{dT_{c}^{<}}{d\delta T}\right|}.
\end{align}
If both temperatures split evenly, the heat-capacity jumps at the two transitions have to be the same. If the upper transition departs more rapidly away from $T_{0}$---which seems to be the case for $\mathrm{Sr}_{2}\mathrm{Ru}\mathrm{O}_{4}$---then the lower heat-capacity jump should be the larger one. This is in sharp disagreement with experiments of Ref.~\cite{Li2020}. 
More freedom is left if two symmetry-distinct order parameters are degenerate by accident and $w \neq 0$ is allowed. Now a negligible second calorimetric jump is possible if $w = - u_{-}$.  However, this condition implies that the second component, i.e. the  $g$-wave order parameter, develops negligible amplitude, i.e. superconductivity prefers to order with $\psi_{d} $ finite but $\psi_{g} \approx 0$.  
\vspace{1em}

\section{Schr\"odinger bound-state problem}
\label{appendixB}
To illustrate the mapping of the local TRSB to the appearance of a bound-state in a Schr\"odinger-type problem we focus on the case with pure $\lambda_{0}$ coupling. Once the system has developed a bulk $d$-wave component $\psi_{d}$ (we assume $\psi_{d}$ to be real) the Ginzburg-Landau equation for the imaginary part $\psi_{g,i} \equiv \im[\psi_{g}]$ reads
\begin{align}\label{}
   \kappa \nabla^{2} \psi_{g,i} = (r_{0} + \delta T + \lambda_{0} \varepsilon^{0})\psi_{g,i} + [(u_{+} + u_{-})/2]\psi_{g,i}^{3}
\end{align}
For simplicity we chose $\kappa \equiv \kappa_{\|} = \kappa_{\perp}$. The parameter $\kappa$ relates to the superconducting coherence length $\xi \sim (\eF/T_{c})a$ via $\kappa \sim \xi^{2} T_{c}$. The cubic term in $\psi_{g,i}$ may be neglected when identifying the transition temperature; this term is relevant to normalizing the bound-state wave-function. The problem can then be cast in the form of a Schr\"odinger equation
\begin{align}\label{}
   \mathcal{H} \Psi = E \Psi
\end{align}
with $\mathcal{H} = -\nabla^{2} + V(\boldsymbol{r})$, $V(\boldsymbol{r}) = \lambda_{0}\varepsilon^{0}/\kappa$, and $E = -(r_{0} + \delta T)/\kappa$. As strain decays with a power-law $b/r$ away from the dislocation, the potential is of Coulomb-type, i.e.\ $V(\boldsymbol{r}) = (\ell/r)\cos(\theta)$, with a characteristic length-scale $\ell^{2} = \kappa/b\lambda_{0}$. For $\kappa_{\|} \neq \kappa_{\perp}$, $\kappa$ should be replaced by the geometric average $(\kappa_{\|} \kappa_{\perp})^{1/2}$ as done in the expression for $\ell$ given in Eq.~\eqref{length}. The above Schr\"odinger problem has a well-defined ground-state energy $E_{0}$, where a bound-state is realized. As temperature is lowered, the effective energy $E$ increases until reaching $E_{0}$. This defines $T = T_{\TRSB}^{\star}$.
\vfill

\end{document}